# Tailoring the morphology of ultrathin bismuth films around percolation for thickness-optimized optical cavities

F. Chacon-Sanchez[1,*], J. Toudert[1], R. Serna[1]

[1]Laser Processing Group, Instituto de Óptica, IO-CSIC, Madrid, Spain

*Corresponding author: fernando.chacon@csic.es

**Abstract:** Ultrathin bismuth films (<10 nm) are emerging candidates for advanced applications in photonics. It has been shown that Bi-based subwavelength optical cavities show outstanding features, including broad tuneable resonances for structural color generation and broadband perfect absorption for light harvesting. While current devices are based on continuous Bi films, integrating ultrathin films around the percolation threshold with a tailored morphology promises to enable more compact devices with a wider tunability range and/or enhanced optical response. In this context, we report the tailoring of the morphology and UV-Vis-NIR effective optical dielectric function ($\varepsilon = \varepsilon_1 + i\varepsilon_2$) of ultrathin Bi films around the percolation threshold. Just below percolation, the films' effective optical permittivity $\varepsilon_1$ differs markedly from that of a continuous film, as it turns from negative to positive while its effective optical loss $\varepsilon_2$ is reduced. We showcase the relevance of such dielectric tunability for the design of enhanced subwavelength optical cavities operating in the NIR and visible regions. In particular, we show that discontinuous near-percolation films enable designing optimized cavities for efficient colour generation with a thickness almost 3 times smaller than what was previously achieved for continuous films.

## 1. Introduction

The search for new materials is at the very core of all materials science research. Following that motivation, new materials with impressive properties have been discovered, like phase change materials, topological insulators or thermoelectric materials, to name a few. These discoveries have resulted in countless improvements in our everyday life.

However, frequently it is sufficient to modify already known materials in a controlled manner to meet the properties needed for given applications. An example of such successful modification is the doping process of silicon that is nowadays the centrepiece of commercial electronics and photovoltaics. In the more specific field of photonics, usual modifications aim at tailoring the morphology of materials at the subwavelength scale to tune their effective optical dielectric function $\varepsilon = \varepsilon_1 + i\varepsilon_2$ to suitable values.

To achieve such tuning, multiple solutions have been devised by engineering metamaterials and metasurfaces. In this approach the selected material is structured at the nanoscale to obtain the desired optical response. [1–4] However, many of the structuration processes used are based on expensive, complex top-down approaches that are difficult to scale for large area processes.

To avoid these drawbacks, the implementation of materials engineered via bottom-up approaches is an appealing option. They enable fabricating nanostructured ultrathin films (thickness < 10 nm) with a tailored morphology and thus, with a tuned effective dielectric function. Such films can be integrated within subwavelength optical cavities to achieve outstanding optical features, such as widespread tunable resonances for color generation or broadband perfect absorption for light harvesting. The so-achieved cavities can be used as compact and lightweight functional surface coatings useful in fields such as anticounterfeiting, photovoltaics, packaging, and more.

The approach of tailoring the optical response of ultrathin films with morphologies ranging from continuous films to nanoparticles has been studied extensively for gold and silver.[5–8] However, the monotonous Drude wavelength dependence of the bulk dielectric function of noble metals sets specific limitations for the obtainable range of values of the effective dielectric function across the electromagnetic spectrum, thus limiting the achievable optical response of the cavities.

On the other hand, the bulk dielectric function of semimetals and other materials based on p-block elements displays a non-spectrally monotonous, non-Drude dielectric function, which shows great potential for optical tailoring.[9] The spectral response of such materials is well exemplified by that of Bismuth (Bi), as can be seen in fig. 1 a). Although a semimetallic elemental material known since ancient times, Bi has only been considered for optical applications recently. Studies aimed at understanding and profiting from its unique optical properties [10] for photonics have flourished during the last 10 years.[11–18] Interestingly, the effective dielectric functions of Bi-based materials reported in the literature differ greatly between them [10], as a result of different nanoscale morphologies due to the different preparation techniques. In particular, pulsed-laser deposited ultrathin films displayed a different effective dielectric function compared with that of thicker films. This difference was ascribed to the progressive loss of continuity in the films as their thickness decreases, as a result of their Volmer-Weber growth. [19]

In these previous works, however, the optical properties of ultrathin Bi films were not explored around the percolation threshold. Herein, we therefore report the tailoring of the morphology and UV-Vis-NIR optical properties of ultrathin bismuth films around this threshold. We show how their effective dielectric function is widely tunable by depositing these ultrathin films with a wide range of morphologies. This, combined with the non-monotonous spectral behaviour of the bulk dielectric function of Bi across the UV-Vis-NIR range, provides an effective tool for the design and fabrication of enhanced Bi-based subwavelength optical cavities. In this context, we explore two designs suited first, for Vis-NIR perfect absorption, and second, for structural color generation. In particular, we highlight the importance of designing the cavities using discontinuous, near-percolation Bi films, in contrast with current devices which are based on continuous ultrathin Bi films.

## 2. Results

We have explored the achievable morphologies for the ultrathin Bi films, by depositing multiple samples with different deposition conditions, but with comparable effective thickness around 10 nm, as observed in fig. 1 c), where we have represented the obtained effective thicknesses for the collection of samples. The films were deposited by Pulsed Laser Deposition (PLD) with an ArF excimer laser, Lambda Physiks LPX-200 laser. Parameters such as the laser energy, the repetition rate, the pressure inside the vacuum chamber, the position of the substrate and the deposition time were fine tuned. Details about the deposition process and parameters can be found in the Supplementary Material, section S1. In addition, the reflectivity at $\lambda = 670\ nm$ was measured in-situ. As a diagram, to illustrate the dependence of the deposition time, which sets the effective deposited thickness on morphology, a typical in-situ reflectivity curve is shown on fig. 1 b). The non-linear increase in reflectance with time indicates that growth occurs following a Volmer-Weber growth mechanism. Three distinct regimes are identified: isolated nanoparticles, near-percolation films and continuous films.

After depositing the films, we characterized their optical properties and thickness by spectroscopic ellipsometry (SE). Employing a Woolam WVASE ellipsometer we measured $\Psi$ and $\Delta$ in the 0.7-5eV range (250-1750nm). To fit the experimental spectra, we created a model and obtained the effective dielectric function $\epsilon$ and the effective thickness $t_{eff}$ of the films. These thicknesses were checked for consistency with AFM (Park Systems XE7, non-contact mode) measurements. More details about the ellipsometric measurements and model can be found in the Supplementary Material, section S2.

As a reference, in fig 1 a) we have represented the dielectric function of bulk Bi, obtained from ref. [8]. We can observe two distinct features, a maximum in $\epsilon_2$ around 0.7 eV and a minimum in $\epsilon_1$ between 1 and 2 eV. This minimum is located within the so-called "metallic window" i.e. where $\epsilon_1$ is negative, like a metal. For thin films, we observed that these two features are still present in the effective dielectric function but vary from film to film, as shown in in figs 1 d) and e). In these figures we have categorized them in three different ranges of thicknesses. This shows that the optical properties can be tuned in a decoupled way from the effective thickness, through the tuning of the film's morphology

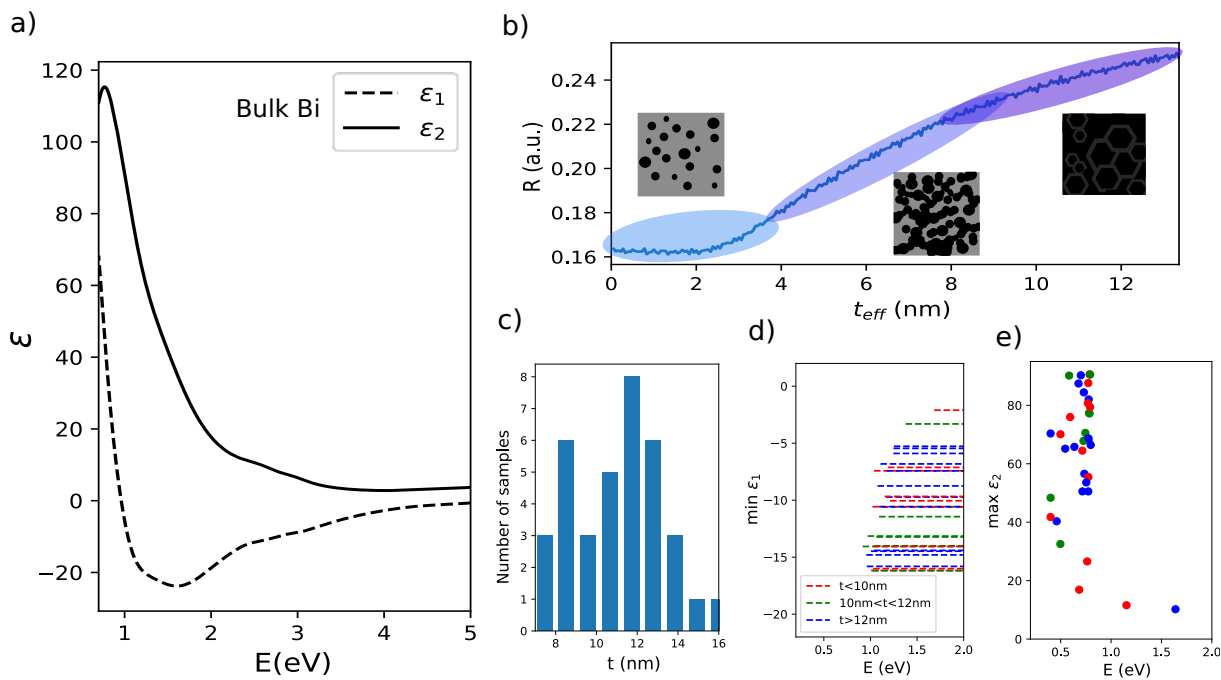

Fig. 1. Exploration of the different morphologies, a) Real ($\epsilon_1$) and imaginary ($\epsilon_2$) dielectric function of bulk Bi in the UV-Vis-NIR range (1750-250 nm). We can observe the maximum on $\epsilon_2$ at low energies, and the minimum on $\epsilon_1$ on the region with $\epsilon_1 < 0$. b) Typical in-situ reflectance measurement during deposition of Bi as a function of the effective thickness, with the three observed regimes differentiated: nanoparticles, near-percolation film and percolated continuous film. c) Effective thickness distribution for the series of deposited films, d) minimum values of $\epsilon_1$. Lines do not indicate the location of said minimum, but rather the energies where $\epsilon_1 < 0$, separated in 3 different stretches of effective thickness, e) maximum values of $\epsilon_2$ along with the energies where said maximum is observed. To facilitate visualization, not all samples in c) are displayed on d) and e).

Fig 1 d) shows the minimum value of the real part of the dielectric function $\epsilon_1$, as a function of the low-energy edge of the metallic window for the different films. We observe a remarkable variation in these minimum values, and an important shift in the edge of the metallic window, from 1 to 1.5 eV approximately (i.e. from 1200 to 800 nm). Figure 1 e) shows the maximum value of the imaginary part of the dielectric function $\epsilon_2$ as a function of its corresponding energy for the different films. The range in variation of $\epsilon_2$ is outstanding, with values ranging from over 90 to less than 20. Furthermore, the energy of this maximum shifts markedly from around 0.7 eV to more than 1.5 eV.

For the interpretation on figs. 1 d) and e), we should remind that Bi nanostructures display plasmon resonances in the UV-Vis-NIR range.[20] Thus, when a Bi film becomes more discontinuous, below percolation, these size-dependent resonances arise and shift toward higher energies. This displaces the lower energy side of the metallic window (fig. 1 d)) and the maximum of $\epsilon_2$ toward higher energies (fig. 1 e)). In addition, as the film becomes more discontinuous, the filling fraction of Bi becomes lower and this smooths out the optical absorption and metallic window, resulting in smaller values of $\epsilon_2$ and values of $\epsilon_1$ closer to zero

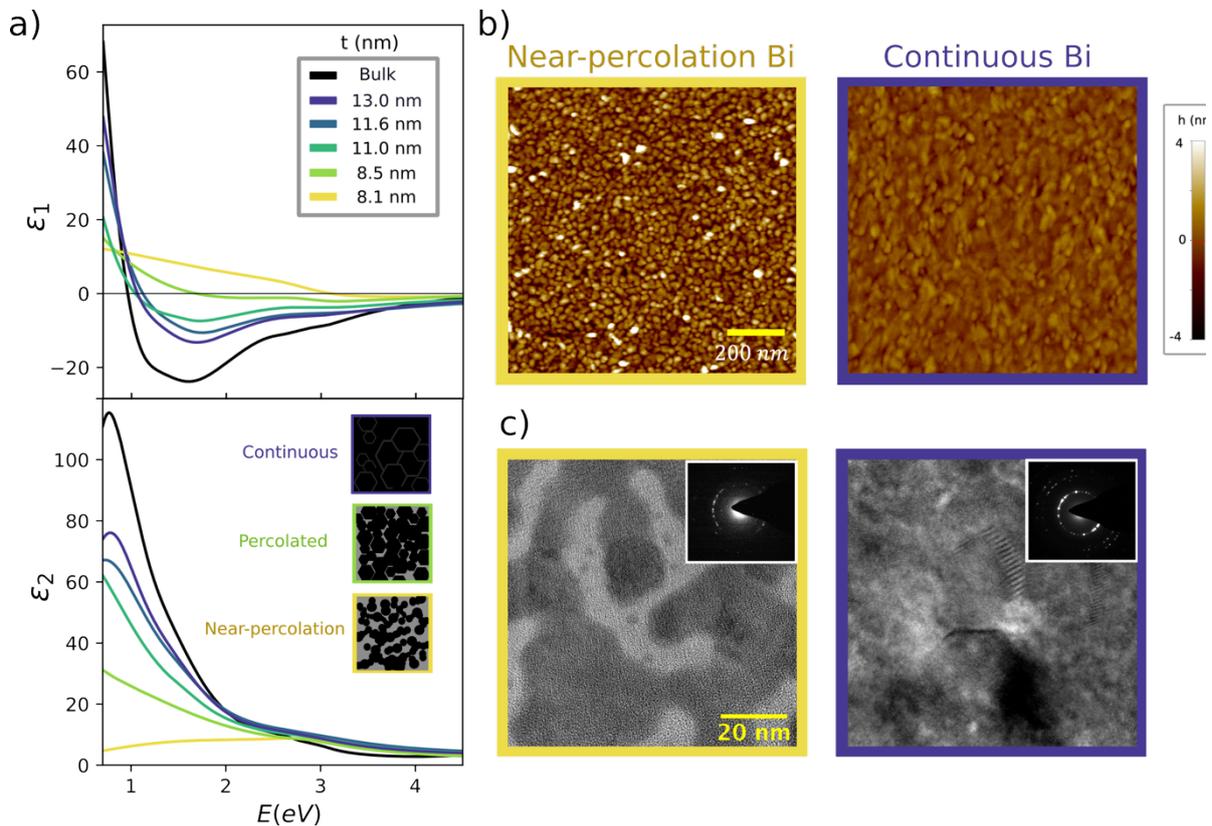

Fig. 2. Advanced characterization of selected ultrathin Bi films. a) Real ($\epsilon_1$) and imaginary ($\epsilon_2$) part of the effective dielectric function, ranging from near-percolation films (yellow) to continuous films (purple), along with bulk Bi, b) AFM images of the near-percolation and continuous films, showing the different morphologies and height differences, c) TEM images of the same films, where we can observe that most of the surface is filled in the continuous film, and that there are empty areas in the near-percolation film. Attached are diffraction diagrams of said films, showing their polycrystalline nature.

On fig. 2 a) we observe the complete effective dielectric functions in the UV-Vis-NIR for a selection of Bi morphologies, ranging from continuous films to near-percolation films just below percolation, as it can be seen for the morphologies of two of the films shown on fig. 2 b) and c). We can clearly differentiate two different regions in the spectrum where each distinct Bi film behaves very differently. The first one corresponds to the UV-Vis ($E > 1.5 eV$, $\lambda < 800 nm$) where we can observe values of $\epsilon_1$ ranging from negative values comparable to those found for metals to positive values like semiconductors. The second one corresponds to the NIR for $E < 1 eV$ ($\lambda > 1240 nm$), where the values of $\epsilon_1$ are positive for all morphologies but the values of $\epsilon_2$ range from low values, more alike common lossy materials, to very high values corresponding with extremely lossy materials. These films are polycrystalline, as can be observed on the corresponding TEM images of fig. 2 c).

This diversity in the values of the effective dielectric function $\epsilon$ illustrates the capacity to tailor the optical properties of ultrathin Bi films depending on the chosen morphology. However, the underlying relevance of this result for photonic applications may not be directly inferred from this figure. In order to better illustrate the importance of this result, we have chosen to computationally study the phenomena of perfect absorption on a Fabry-Perot (F-P) cavity, as perfect absorption is a very sought-after result, that has been studied through many different approaches [17,21–23] In addition, a F-P cavity is a very simple structure that can provide great results on this pursuit [13]. We considered structures similar to the so-called MIMI cavities, in which an ultrathin Bi film is sandwiched between Si layers, the structure standing on top of a thick Al layer. We calculated the reflectance of the cavity using different effective dielectric functions for the Bi film, taken from fig. 2, to evaluate the effect of this layer's morphology – from near-percolation to continuous on the cavity's response. We have chosen aluminium (Al) as the metallic back mirror. We selected silicon (Si) as the dielectric, as for the studied wavelengths it has very low losses ($k < 0.02$, $\epsilon_2 < 0.2$), and for the thicknesses considered the base absorption is very low. Additionally, we chose typical thicknesses for the two layers of Si to obtain first order destructive interferences in the NIR and Vis, with 60 and 30 nm, respectively, and 9 nm for all the ultrathin Bi, as can be observed in figure 2 a). We have fixed the thickness of Bi to reduce the number of variable parameters within the structure and better illustrate the difference in the dielectric function, but the results with the actual measured values don't differ in a relevant way, as can be observed in fig. S3.

On fig 3 b) we observe the resulting reflectance of this structure at normal incidence. We observe three clear minima in the spectrum, where the first two depend on the implemented morphology and the third one doesn't. This last minimum is placed exactly at the edge of the electronic transitions of Si and hence it will not be taken into consideration. The other two, however, show a great dependence on the morphology of the ultrathin Bi film. Insets in Fig 3 b) shows a detailed representation of each one. If we focus on the first one and analyse it along the dielectric functions shown in fig 2 a) (detailed in fig. S2), we conclude that at the first dip in reflectivity, higher values of the imaginary part of the effective dielectric function $\epsilon_2$ of the ultrathin Bi film yield a higher absorption by the cavity in most cases. The only exception is that of a hypothetic bulk-like Bi ultrathin film where $\epsilon_2$ values are too high and we observe the phenomenon known as overcoupling [21]. From this and the absorption in individual layers of the cavity shown in fig. 3 c) in the cases of the archetypal continuous film and near-percolation films we can deduce that, contrary to general intuition, higher optical impedance mismatch does not result in higher reflectivity, but rather in higher absorption.

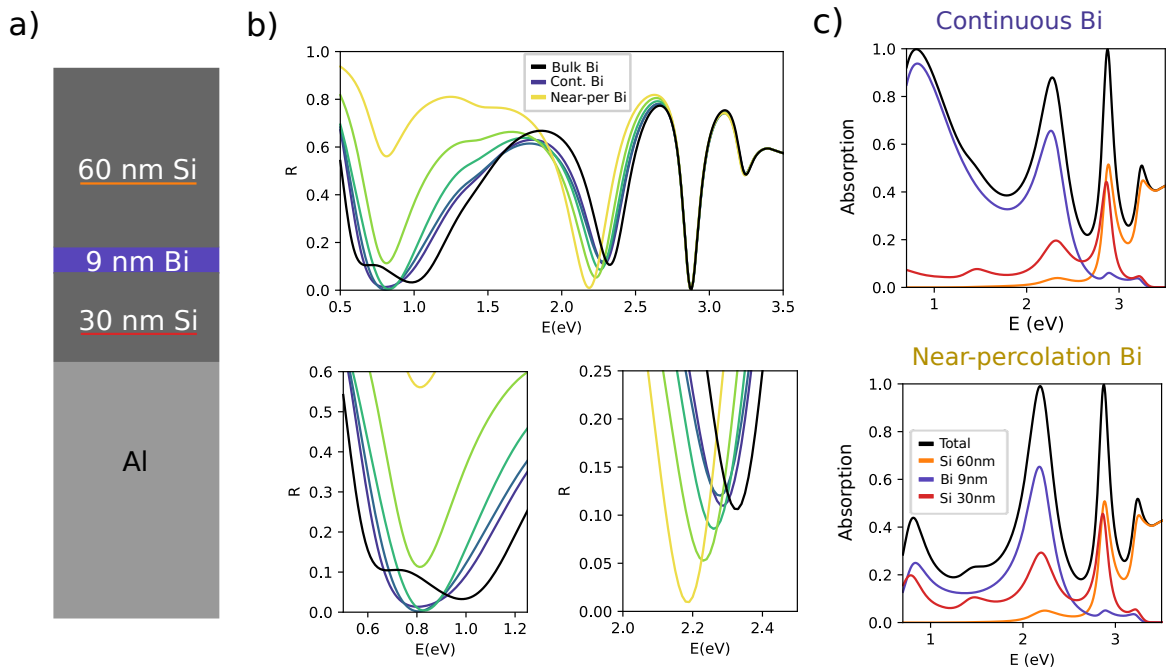

Fig. 3. Tailoring perfect absorption of subwavelength optical cavities by tuning the morphology of the ultrathin Bi film. a) Schematic of the simulated structure, with the chosen values for each thickness. Silicon thicknesses are typical values to generate destructive interferences in the IR and visible, respectively, and Bi thickness is a generic value in the range of the studied morphologies, b) simulated reflectivity at normal incidence in the Vis-NIR range, with detailed graphs of the first two minima, c) optical absorption in the individual layers of the cavity, when considering a continuous Bi film (purple on b) and a near-percolation Bi film (yellow on b).

On the other hand, for the second dip in reflectivity, we observe the opposite behaviour, as less continuous films result in higher absorption. This phenomenon can be explained by analysing the absorption in the individual layers of the cavity shown in fig 3 c). They show that absorption in the Bi film saturates around 0.65 in the case of both the continuous and near-percolation films. In the meantime, extra absorption is observed inside the bottom Si layer in the case of the near-percolation film. This is due to the fact that this film shows values of $\epsilon_1 > 0$, i.e., it behaves optically as a lossy dielectric and thus does not reflect light as efficiently as continuous films, which behave optically as metals ($\epsilon_1 < 0$). Therefore, the near-percolation Bi film reflects a smaller amount of incident light than continuous ones before it enters the bottom cavity, hence helping light to be trapped in the cavity and be better absorbed.

To sum it up, we have computationally studied a generic F-P cavity and observed deep sub-wavelength absorption in Bi, with each extreme morphology (continuous vs near-percolation) showing a part of the electromagnetic range more suitable to obtain perfect absorption (NIR vs Vis). This can be traced to the lower effective intensity of the interband transitions in the near-percolation nanostructures in the IR and to the higher metallicity of the smooth ultrathin nanostructures in the Vis.

To further inspect the usefulness of the tailored ultrathin Bi films, we studied their application for sustainable structural coloring. For this purpose, we followed the Semimetal-Substrate Cavity (SSC) approach presented in ref 24. This SSC approach implements industrial and everyday materials directly as the back mirror to generate structural colors. This provides a simple, cost-effective alternative to the omnipresent metallic back mirrors. This is combined with MIMI-like cavities and semimetals to generate broad destructive interferences that result in vivid reflective structural colors.

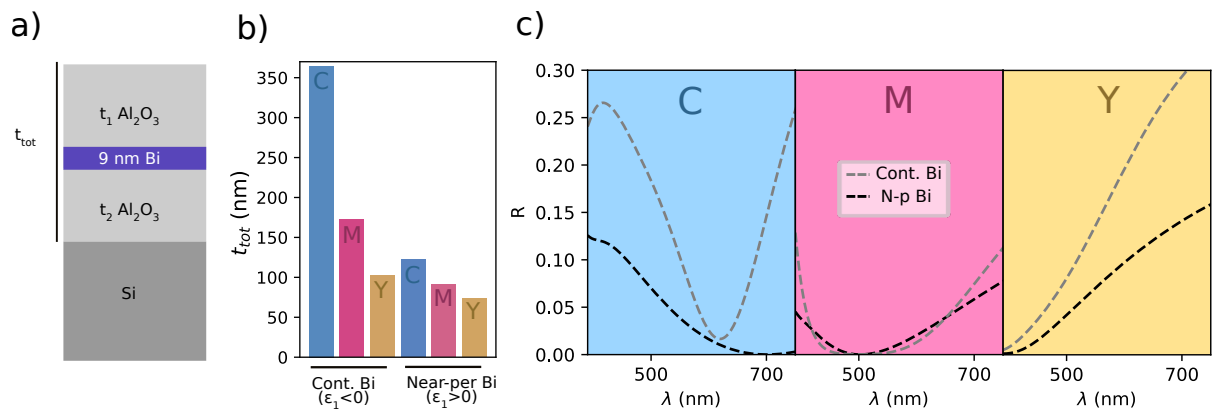

Fig. 4. Color generation with different ultrathin Bi film morphologies in a SSC cavity. a) Schematic of the studied SSC structure, following the approach described in ref. 24. b) Optimized total thickness of the structure to obtain a Cyan-Magenta-Yellow (CMY) colorbase on SSC cavities using Si as the substrate in the cases of cavities based on continuous and near-percolation ultrathin Bi films. Each bar is colored with the obtained color. c) Simulated reflectivity for each CMY colorbase for both types of Bi films.

On fig. 4 we computationally studied the generation of the Cyan, Magenta and Yellow (CMY) colorbase for two different SSCs, one including a continuous Bi film with a dielectric function similar to the one of bulk Bi (as in ref. 24), and the other the same near-percolation film considered above. The general structure considered is pictured in fig. 4 a), with the thickness of each $Al_2O_3$ dielectric layer optimized for each color/film combination and choosing Si as the substrate. On fig. 4 b) we have represented the minimal total thickness of the structure needed to obtain each optimized color for the two cavities, with the color of each bar representing the obtained CIE color. Details about the needed dielectric thickness for each individual dielectric layer can be found on figure S4. We observe that the obtained colors are similar for the different films, with high accuracy and resemblance to the target colours. However, the total dielectric thickness needed differs greatly, given that a total cavity thickness up to 3 times smaller can be implemented to obtain accurate colors when choosing a near-percolation Bi film. The obtained reflectance for each color/film is displayed in fig. 4 c).

This can be readily explained by considering that for the SSC with the continuous film, this film displays a metallic behaviour and effectively the system is comprised of two different but coupled cavities, whereas for the non-metallic near-percolation structure it acts effectively as a lossy single cavity, hence requiring around half the thickness to obtain perfect absorption at a desired wavelength.

In conclusion, we have demonstrated the potential of the morphology tailoring on ultrathin Bi films, displaying an outstanding range with very different behaviour in different parts of the electromagnetic spectrum. As a result, we have been able to access an outstanding range of optical properties that has in turn allowed us to choose a specific morphology depending on the desired application. To illustrate that, we have shown how different morphologies are more suitable for different applications, from perfect absorption in the NIR and Vis to structural color generation, with cavity thicknesses up to 3 times smaller than previously reported. This morphology tailoring can prove relevant for nanophotonic applications, including anticounterfeiting, and sensing with thinner and more compact devices.

**Acknowledgements**

This work has been partly funded by the national research grants SLIM-2P (PID2024-156974OB-C21) and ALPHOMENA (PID2021-123190OB-I00) funded by MICIU/AEI/ 10.13039/501100011033/FEDER,UE. Authors gratefully acknowledge the use of instrumentation as well as the technical advice provided by the National Facility ELECMI ICTS, node "Laboratorio de Microscopías Avanzadas" at University of Zaragoza.

The following article has been submitted to *Applied Physics Letter*s. After it is published, it will be found at Link

# References

[1] C. Ruiz de Galarreta, I. Sinev, A.M. Alexeev, P. Trofimov, K. Ladutenko, S. Garcia-Cuevas Carrillo, E. Gemo, A. Baldycheva, J. Bertolotti, and C. David Wright, "Reconfigurable multilevel control of hybrid all-dielectric phase-change metasurfaces," Optica **7**(5), 476 (2020).

[2] L. Feng, P. Huo, Y. Liang, T. Xu, L. Feng, P. Huo, Y. Liang, and T. Xu, "Photonic Metamaterial Absorbers: Morphology Engineering and Interdisciplinary Applications," Advanced Materials **32**(27), 1903787 (2020).

[3] J. Olson, A. Manjavacas, L. Liu, W.S. Chang, B. Foerster, N.S. King, M.W. Knight, P. Nordlander, N.J. Halas, and S. Link, "Vivid, full-color aluminum plasmonic pixels," Proc Natl Acad Sci U S A **111**(40), 14348–14353 (2014).

[4] R.A. Maniyara, D. Rodrigo, R. Yu, J. Canet-Ferrer, D.S. Ghosh, R. Yongsunthon, D.E. Baker, A. Rezikyan, F.J. García de Abajo, and V. Pruneri, "Tunable plasmons in ultrathin metal films," Nat Photonics **13**(5), 328–333 (2019).

[5] J. Kim, H. Oh, M. Seo, and M. Lee, "Generation of Reflection Colors from Metal-Insulator-Metal Cavity Structure Enabled by Thickness-Dependent Refractive Indices of Metal Thin Film," ACS Photonics **6**(9), 2342–2349 (2019).

[6] M. Hövel, B. Gompf, and M. Dressel, "Dielectric properties of ultrathin metal films around the percolation threshold," Phys Rev B **81**(3), 035402 (2010).

[7] M. Novotný, P. Fitl, S.A. Irimiciuc, J. Bulíř, J. More-Chevalier, L. Fekete, P. Hruška, S. Chertopalov, M. Vrňata, and J. Lančok, "In situ monitoring of electrical resistivity and plasma during pulsed laser deposition growth of ultra-thin silver films," J Appl Phys **130**(8), (2021).

[8] J.C.G. De Sande, R. Serna, J. Gonzalo, C.N. Afonso, D.E. Hole, and A. Naudon, "Refractive index of Ag nanocrystals composite films in the neighborhood of the surface plasmon resonance," J Appl Phys **91**(3), 1536–1541 (2002).

[9] J. Toudert, and R. Serna, "Interband transitions in semi-metals, semiconductors, and topological insulators: a new driving force for plasmonics and nanophotonics [Invited]," Opt Mater Express **7**(7), 2299 (2017).

[10] J. Toudert, R. Serna, I. Camps, J. Wojcik, P. Mascher, E. Rebollar, and T.A. Ezquerra, "Unveiling the Far Infrared-to-Ultraviolet Optical Properties of Bismuth for Applications in Plasmonics and Nanophotonics," Journal of Physical Chemistry C **121**(6), 3511–3521 (2017).

[11] M. Garcia-Pardo, E. Nieto-Pinero, A.K. Petford-Long, R. Serna, and J. Toudert, "Active analog tuning of the phase of light in the visible regime by bismuth-based metamaterials," Nanophotonics **9**(4), 885–896 (2020).

[12] M. Alvarez-Alegria, J. Siegel, M. Garcia-Pardo, F. Cabello, J. Toudert, E. Haro-Poniatowski, and R. Serna, "Nanosecond Laser Switching of Phase-Change Random Metasurfaces with Tunable ON-State," Adv Opt Mater **10**(3), (2022).

[13] F. Chacon-Sanchez, C.R. de Galarreta, E. Nieto-Pinero, M. Garcia-Pardo, E. Garcia-Tabares, N. Ramos, M. Castillo, M. Lopez-Garcia, J. Siegel, J. Toudert, C.D. Wright, and R. Serna, "Building Conventional Metasurfaces with Unconventional Interband Plasmonics: A Versatile Route for Sustainable Structural Color Generation Based on Bismuth," Adv Opt Mater **12**(10), 2302130 (2024).

[14] I. Ozbay, A. Ghobadi, B. Butun, and G. Turhan-Sayan, "Bismuth plasmonics for extraordinary light absorption in deep sub-wavelength geometries," Opt Lett **45**(3), 686 (2020).

[15] D. Zhou, J. Zhang, L. Li, C. Tan, Z. Zhang, Y. Sun, L. Zhou, N. Dai, J. Chu, and J. Hao, "Subwavelength broadband light-harvesting metacoating for infrared camouflage and anti-counterfeiting empowered by inverse design," Materials Today Physics **50**, 101614 (2025).

[16] J. Hao, D. Zhou, C. Tan, Q. Qiu, Z. Zhang, Y. Sun, J. Zhang, L. Li, L. Zhou, N. Dai, and J. Chu, "Semimetal-dielectric-metal metasurface for infrared camouflage with high-performance energy dissipation in non-atmospheric transparency window," Nanophotonics **14**(8), 1101–1111 (2025).

[17] J. Toudert, R. Serna, M.G. Pardo, N. Ramos, R.J. Peláez, and B. Maté, "Mid-to-far infrared tunable perfect absorption by a sub - λ/100 nanofilm in a fractal phasor resonant cavity," Opt Express **26**(26), 34043 (2018).

[18] A. Cuadrado, J. Toudert, and R. Serna, "Polaritonic-to-Plasmonic Transition in Optically Resonant Bismuth Nanospheres for High-Contrast Switchable Ultraviolet Meta-Filters," IEEE Photonics J **8**(3), (2016).

[19] J. Toudert, R. Serna, C. Deeb, and E. Rebollar, "Optical properties of bismuth nanostructures towards the ultrathin film regime," Opt Mater Express **9**(7), 2924 (2019).

[20] J. Toudert, R. Serna, and M. Jiménez De Castro, "Exploring the Optical Potential of Nano-Bismuth: Tunable Surface Plasmon Resonances in the Near Ultraviolet-to-Near Infrared Range," Journal of Physical Chemistry C **116**(38), 20530–20539 (2012).

[21] W.J. Chang, Z. Sakotic, A. Ware, A.M. Green, B.J. Roman, K. Kim, T.M. Truskett, D. Wasserman, and D.J. Milliron, "Wavelength Tunable Infrared Perfect Absorption in Plasmonic Nanocrystal Monolayers," ACS Nano **18**(1), 972–982 (2024).

[22] M.K. Hedayati, M. Javaherirahim, B. Mozooni, R. Abdelaziz, A. Tavassolizadeh, V.S.K. Chakravadhanula, V. Zaporojtchenko, T. Strunkus, F. Faupel, and M. Elbahri, "Design of a Perfect Black Absorber at Visible Frequencies Using Plasmonic Metamaterials," Advanced Materials **23**(45), 5410–5414 (2011).

[23] G. Gomez-Muñoz, R. Benítez-Fernández, G. Godoy-Perez, F. Cabello, M. Garcia-Pardo, D. Caudevilla, J. Gonzalo, J. Solis, M. Garcia-Lechuga, J. Olea, D. Pastor, and J. Siegel, "Tuning spike-like morphologies in Silicon by sustainable fs-laser processing in air for enhanced light absorption," Appl Surf Sci, 161967 (2024).

[24] F. Chacon-Sanchez, and R. Serna, "Semimetal/Substrate Cavities Enabling Industrial Materials for Structural Coloring," ACS Applied Optical Materials **3**(3), 727–736 (2025).

# Supplementary Material

S1. Details about deposition of Bi nanostructures.

The ultrathin Bi films were deposited on crystalline silicon substrates with their native (2 nm thick) oxide layer using an in-house Pulsed Laser Deposition (PLD) system. Deposition parameters on the collection of samples studied were between 2-4 J/cm$^2$ for the laser energy density per pulse, with a repetition rate between 5 and 20 Hz. . The chamber background pressure was varied between $2.5 \times 10^{-6}$ to $1.5 \times 10^{-7}$ mbar. Lastly, the deposition time was changed between 60 and 120 s.

S2. Details about the ellipsometric measurements.

Optical characterization of the studied films was performed via Spectroscopic Ellipsometry, employing a Woollam WVASE ellipsometer in the 0.7-5 eV range (250-1750 nm). All nanostructures were modelled via a combination of Lorentz oscillators, with a resulting Medium Square Error (MSE) lower than 10 for all cases. Roughness was not considered as a different layer and it's included in the effective optical properties of each film. On the following table the ellipsometric model employed for the 11.6 nm bi film (blue line on fig 2 on the main text), with a very good fit with the measurements, as can be seen on figure S1, yielding a MSE of 2.3.

| Oscillator | A | En | Br |
|---|---|---|---|
| Lorentz | 56.9 | 0.96 | 1.40 |
| Lorentz | 3.6 | 2.93 | 1.79 |
| Lorentz | 2.4 | 7.03 | 10.00 |
| Lorentz | 5.3 | 1.59 | 0.84 |

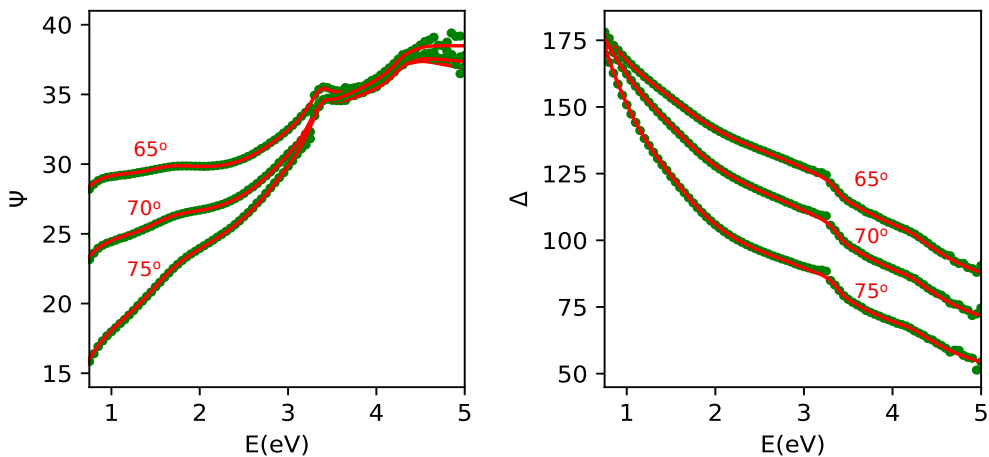

Fig. S1. Measured and fitted ellipsometric parameters psi and delta for the 11.6 nm continuous film.

## S3. Details about perfect absorption structure.

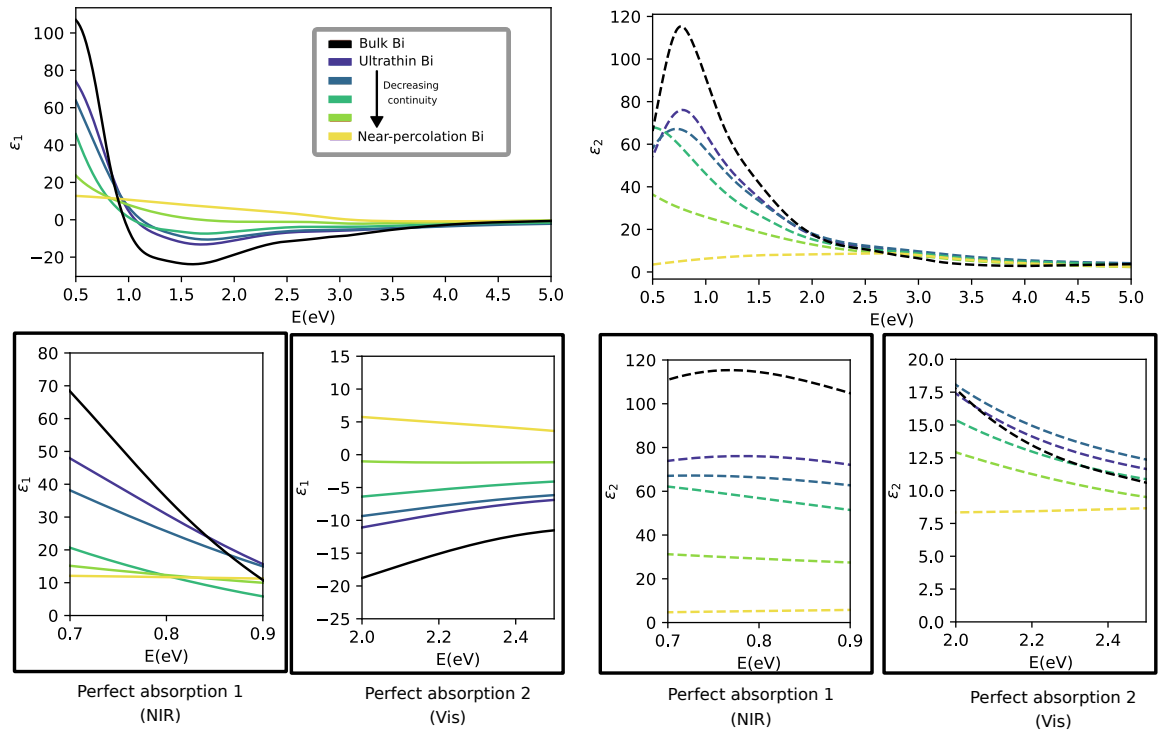

Fig. S2. Real and imaginary part of the dielectric function of selected Bi nanostructures. The top panels show the real ($\varepsilon_1$) and imaginary ($\varepsilon_2$) dielectric function in the full studied range 0.7-5 eV (analogous to fig. 2 a)). In the bottom panels we show the detail of their values around the energies where perfect absorption is observed in fig.3 in each case (0.7 to 0.9 eV region for perfect absorption around 0.8 eV, and 2.0 to 2.5 eV region for perfect absorption around 2.2 eV).

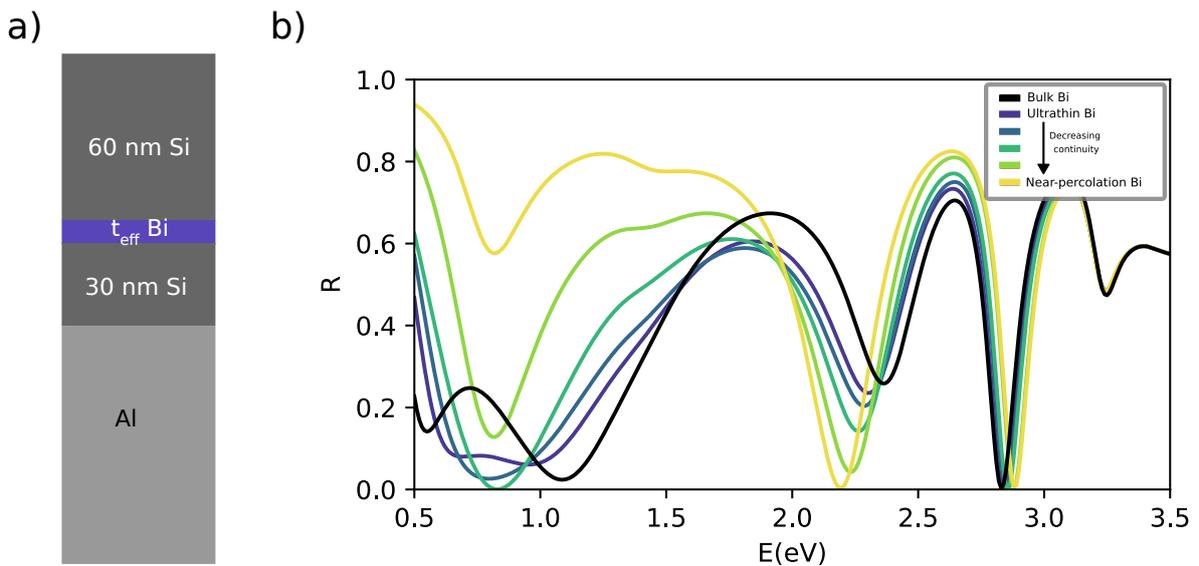

Fig. S3. Reflectivity for the perfect absorber with real thicknesses. a) Schematic of the considered structure. Its only difference with the one considered for figure 3 in the main text is the Bi thickness, which is now taken as the effective thicknesses of each film. b) Reflectivity of said structure in the Vis-NIR. No remarkable differences are observed in the shape of the spectra compared to Figure 3, except for some overcoupling observed for bulk-like films and a slight displacement on the energy of the minima between films.

## S4. Details about structural coloring with Bi ultrathin films.

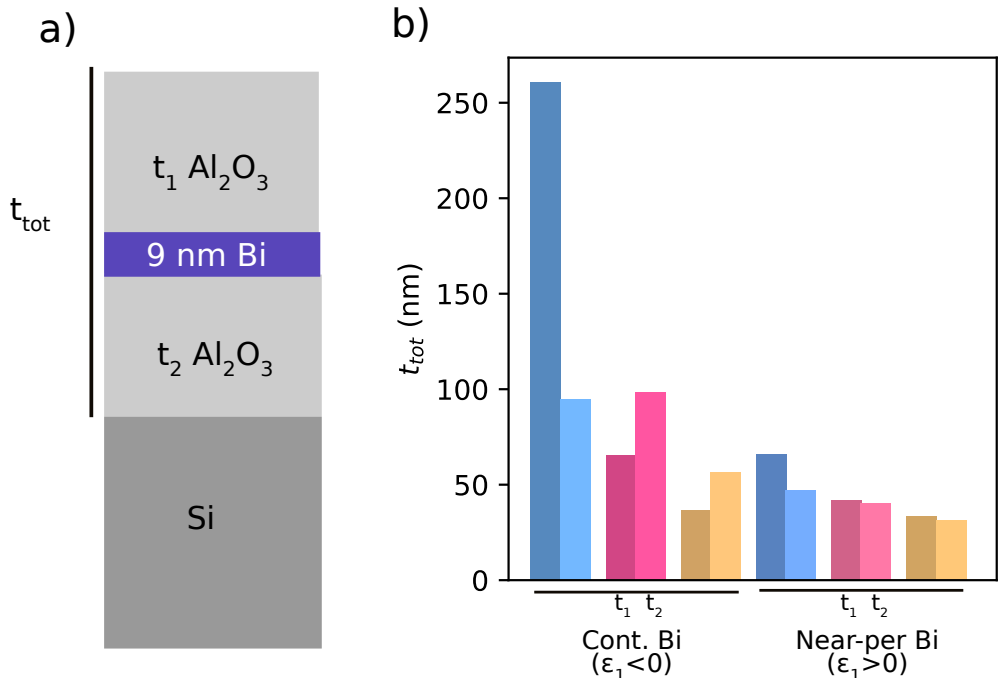

Figure S4. Thickness of each dielectric layer for SSCs. a) General schematic of the optimized structures. b) Optimized thickness of upper($t_1$) and lower($t_2$) $Al_2O_3$ layers for the optimized CMY colorbase for continuous and near-percolation Bi films